\documentclass[12pt]{article}

\usepackage{amssymb}
\usepackage{amsmath}
\usepackage{amsfonts}
\usepackage{graphicx}
\usepackage{mathrsfs}
\usepackage{comment}
\usepackage{cite}
\usepackage{authblk}
\usepackage{array}
\usepackage{subcaption}

\newcommand{\Slash}[1]{{\ooalign{\hfil#1\hfil\crcr\raise.167ex\hbox{/}}}}

\newcommand{\beq}{\begin{equation}}  \newcommand{\eeq}{\end{equation}}
\newcommand{\bef}{\begin{figure}}  \newcommand{\eef}{\end{figure}}
\newcommand{\bec}{\begin{center}}  \newcommand{\eec}{\end{center}}
  
\newcommand{\laq}[1]{\label{eq:#1}}  

\newcommand{\Eq}[1]{Eq.(\ref{eq:#1})}

\newcommand{\eq}[1]{(\ref{eq:#1})}

\newcommand{\ab}[1]{\left|{#1}\right|}
\newcommand{\vev}[1]{ \left\langle {#1} \right\rangle }

\def\({\left(}
\def\){\right)}

\def\O{\mathcal{O}}

\newcommand{\AND}{~{\rm and}~}
\newcommand{\EV}{ {\rm \, eV} }

\newcommand{\MEV}{ {\rm \, MeV} }
\newcommand{\GEV}{ {\rm \, GeV} }

\def\o{\over}
\def\a{\alpha}

\def\d{\delta}
\def\e{\epsilon}
\def\f{\phi}
\def\g{\gamma}

\def\k{\kappa}

\def\p{\psi}

\def\s{\sigma}
\def\t{\tau}

\def\x{\xi}

\def\L{\Lambda}

\def\tl{\tilde}
\def\*{\dagger}

\begin{document}
\begin{titlepage}
\begin{center}

\hfill   TU-1023

\vspace{1.0cm}

{\Large\bf Cosmic Axion Force}
\vspace{1.5cm}

{\bf Dongok Kim$^{1,2}$, Younggeun Kim$^{1,2}$, Yannis K. Semertzidis$^{2,1}$, \\
Yun Chang Shin$^{2}$,  Wen Yin$^{3,4}$}

\vspace{1.5cm}
{\em 
$^{1}$Department of Physics, Korea Advanced Institute of Science and Technology, Daejeon 34141 Republic of Korea,\\
$^{2}$Center for Axion and Precision Physics Research, Institute for Basic Science, Daejeon 34051 Republic of Korea,\\
$^{3}${Department of Physics, Tohoku University,  Sendai, Miyagi 980-8578, Japan \\} 
$^{4}${Department of Physics, Faculty of Science, 
The University of Tokyo,  Bunkyo-ku, Tokyo 113-0033, Japan\\}
}

\vspace{1.0cm}
\abstract{
Nambu-Goldstone bosons, or axions, may be ubiquitous. 
Some of the axions  may have small masses and thus serve as mediators of long-range forces. 
In this paper, we study the force mediated by an extremely light axion, $\phi$, between the visible sector and the dark sector, where dark matter lives.  
Since nature does not preserve the CP symmetry, the coupling between dark matter and $\phi$ is generically CP-violating.
In this case, the induced force is extremely long-range and behaves as an effective magnetic field. If the force acts on electrons or nucleons, the spins of them on Earth  precess around a fixed direction towards
the galactic center. 
This provides an experimental opportunity for $\phi$ with mass, $m_\phi$, and decay constant, $f_\phi$, satisfying $m_\phi\lesssim 10^{-25}\,$ eV, $f_\phi\lesssim 10^{14}\,$GeV if the daily modulation of the effective magnetic field signals in magnetometers is measured by using the coherent averaging method.
The effective magnetic field induced by an axionic compact object, such as an axion domain wall, is also discussed. 
}

\end{center}
\end{titlepage}

\setcounter{footnote}{0}
\section{Introduction}

Despite the evidence of CP violation, the strong sector has a very good CP symmetric structure, which is unnatural and dubbed as a strong CP problem. 
This problem may be solved by the existence of a QCD axion~\cite{Peccei:1977hh,Peccei:1977ur,Weinberg:1977ma,Wilczek:1977pj}, 
which makes the QCD sector of the standard model (SM) settle into an almost CP-conserving vacuum.

Interestingly, string or M-theory may predict axiverse~\cite{Witten:1984dg, Svrcek:2006yi,Conlon:2006tq,Arvanitaki:2009fg}, in which huge amounts of axions exist. 
The masses are generated by non-perturbative effects and thus spread over a wide range. Some of the string axions may be heavy while some may be very light. 
Generic axions, or pseudo-Nambu-Goldstone bosons, have also been discussed widely~\cite{Acharya:2010zx, Higaki:2011me, Cicoli:2012sz, Daido:2016tsj, Agrawal:2017ksf, Graham:2018jyp, Guth:2018hsa,  Demirtas:2018akl, Co:2018phi, Ho:2019ayl, Matsui:2020wfx,Yanagida:2019evh, Takahashi:2019pqf, Arvanitaki:2019rax, Marsh:2019bjr, Nakagawa:2020eeg, Kitano:2021fdl}, especially in the context of the dark matter (DM). 
Other than the dominant DM, some axion may form topological defects like cosmic strings or domain walls~\cite{Kibble:1976sj, Kibble:1980mv, Takahashi:2020tqv}, and some may become the dark radiation~\cite{Cicoli:2012aq,Higaki:2012ar,Conlon:2013isa,Hebecker:2014gka, Dror:2021nyr, Jaeckel:2021gah}. 
Phenomenologically, a hint of the isotropic cosmic birefringence of cosmic microwave background polarization was reported~\cite{Minami:2020odp}. 
The birefringence can be explained by the existence of a very light axion(s) or axionic topological defects coupled to photons~\cite{Minami:2020odp,Fujita:2020ecn,Takahashi:2020tqv, Mehta:2021pwf, Nakagawa:2021nme}
(see also \cite{Carroll:1989vb, Carroll:1991zs, Harari:1992ea,Carroll:1998zi,Lue:1998mq,Pospelov:2008gg,Fedderke:2019ajk,Agrawal:2019lkr,Jain:2021shf}).  See
Refs.\,\cite{Jaeckel:2010ni,Ringwald:2012hr,Arias:2012az,Graham:2015ouw,Marsh:2015xka,Irastorza:2018dyq, DiLuzio:2020wdo} for reviews.

The axion can also play the role of long-range force contributing to the monopole-monopole, monopole-dipole and dipole-dipole type interactions between the visible sector particles~\cite{Moody:1984ba, Pospelov:1997uv}. (See also \cite{Adelberger:2009zz, Jaeckel:2010ni, Safronova:2017xyt}.) To have a monopole source we need a CP-violation between the axion and visible particles. Although the CP feature of the SM highly suppresses the monopole coupling, one may get constraints comparable to the astronomical one~\cite{Raffelt:2012sp}. Furthermore, the monopole-dipole interaction may provide strong evidence of the presence of the QCD axion in the ARIADNE experiment~\cite{Arvanitaki:2014dfa,Geraci:2017bmq}. In general, the force becomes longer range if the axion mass is lighter.

 In any case, a dark sector must exist to explain the missing mass of the Universe. 
 Here we notice that, unlike the unnatural QCD sector, the dark sector may be generically CP-violating.
For instance, as we show in  Appendix \ref{ap:1}, in a dark QCD model, the dark nucleon provides a monopole force if we do not tune the strong CP phase. 
With several axions as in the axiverse, the vacuum of the axion potential can be CP-violating, and then the axion DM emits the monopole force. 
Also, axionic topological defects generally emit the monopole force.

\begin{figure}[t!]
\centering
\includegraphics[width=1\textwidth]{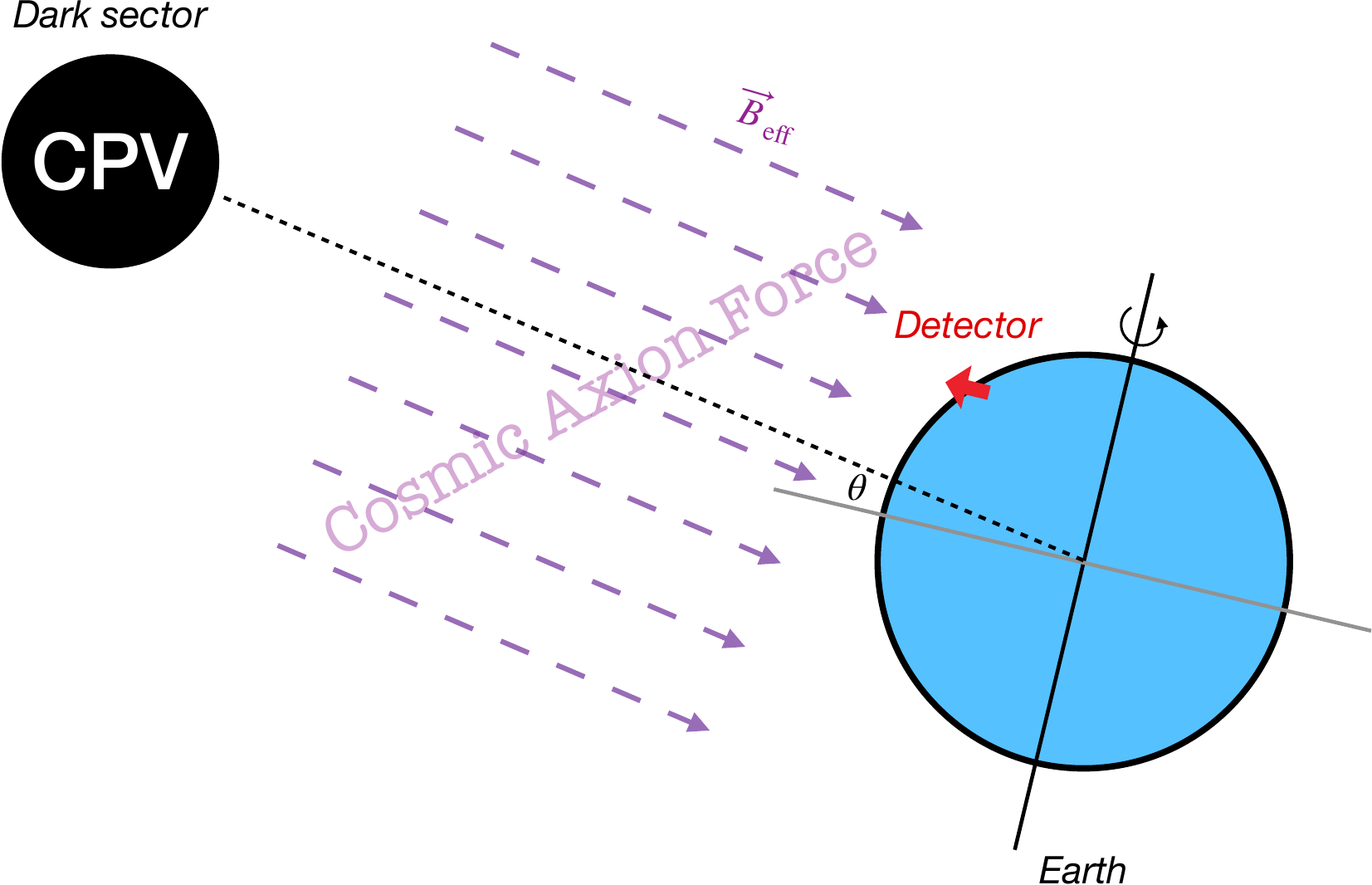}
\caption{A conceptual diagram of the detection for the cosmic axion force with a magnetometer. 
The long-range force emitted from a CP-violating dark sector, e.g. DM or compact objects, (black) behaves like a constant magnetic field (purple dashed lines), that cannot be shielded as an ordinary magnetic field. 
Since the Earth (blue) is rotating along its axis, signals from the cosmic axion force measured by a local detector (red) have a daily modulation depending on the declination angle $\theta$. }\label{fig:scheme}
\end{figure}

In this paper, therefore, we study the possibility that an ultra-light axion plays the role of cosmically long-range force between the CP-violating dark sector and the visible sector particles. We call it a cosmic axion force.
A dark sector particle or a compact object emits the cosmic axion force with a ``monopole potential" $\propto 1/r$.  
We point out that via the monopole-dipole interaction, the cosmic axion force behaves as an effective magnetic field which induces the spin precessions of $\f$-coupled SM fermions all over the Earth around a fixed direction (see Fig.\ref{fig:scheme}). This axion-induced magnetic field remains intact when the ordinary magnetic fields are shielded. 
Such precession can be identified by 
 carefully measuring the daily 
 modulation of the effective magnetic field by magnetometers.

 Magnetometers are popularly used in direct detections of the axion DM in the experiments of ABRACADABRA~\cite{Kahn:2016aff,Ouellet:2018beu}, and
 CASPEr~\cite{Graham:2013gfa, Budker:2013hfa, JacksonKimball:2017elr, Garcon:2017ixh}. 
 In these cases, the induced magnetic field is along the direction of the DM velocity and the strength is time-varying. 
In contrast, the magnetic field induced by the cosmic axion force is towards the direction of the denser place of the source, e.g. the galactic center  for the DM, 
and the strength is almost constant in time, and, therefore, distinguishable from the direct detection of the DM. 
 Compact object can be searched for in the magnetometers of GNOME ~\cite{Pustelny:2013rza,Afach:2018eze,Afach:2021pfd}. The induced effective magnetic field is also time-varying.

This paper is organized as follows. 
In the next section we give the theoretical background on the cosmic axion force emitted from DM and compact object, and estimate the effective magnetic field for the spin precession. In  Sec.\,\ref{sec:3}, we discuss the experimental opportunity. The last section is devoted to the conclusions.

\section{CP-violating axion and axion force}

An axion, $\f$ which can be seen as a pseudo-Nambu-Goldstone boson, 
is a CP-odd scalar. Thus with CP symmetry and the shift symmetry, under which, respectively, axion transforms \beq \phi\to -\phi \AND \f \to \f +\a \laq{shift},\eeq the $\f$-couplings to particles are controlled.
Here $\a$ is a real arbitrary number. 
However CP is violated in nature. 
This means, in general, the axion couples to particles, such as those in the dark sector, via CP-violating interaction (See Appendix.~\ref{ap:1} for concrete examples).

The exception is the couplings to the SM fermions. The coupling, especially to the nucleon, is almost CP-symmetric since CP symmetry miraculously exists in the QCD sector: the nucleon EDM has not been observed. 
There are also severe constraints in the fifth force searches~\cite{Schlamminger:2007ht,Stadnik:2016zkf}, which not only constrains the CP-violating coupling to nucleon but also electron. 

In this section, we study the axion mediated long-range force between the CP-violating dark sector and CP-preserving SM sector (by neglecting the CKM induced effect for simplicity unless otherwise stated). 
We emphasize that we will not take $\f$ as the QCD axion throughout this paper, but we consider it as a generic axion particle that may arise from string/M-theory. 
The potential (around the minimum) is assumed to be 
\beq
V\supset \frac{m_\phi^2}{2}\phi^2.
\eeq
This term violates the shift symmetry \eq{shift}. 
We write this term since we would like to clarify the viable range for $\f$ mass in our scenario. 
As we will see with $m_\f\to 0$, our mechanism also works. 

\subsection{Cosmic axion force from CP-violating dark sector}
Let us estimate the axion force from a CP-violating source, $J(t,\vec{x})$, in the dark sector.
The equation of motion of $\d\f=\f-\vev{\f}$ with $\vev{\f}$ being the vacuum expectation value (VEV), in flat spacetime, is given by 
\beq
\laq{eom}
(\partial_t^2- \nabla^2) \d \phi (t, \vec{ x})= -m^2_\f \d\phi (t, \vec{x})-J(t,\vec{x}), 
\eeq
where $m_\f$ is the mass of $\f$. 
We parametrize 
\beq
J= \epsilon C[t]\frac{\rho_{\rm DS}(\vec{x})}{f_\f},
\eeq
where $\e$ quantifies the CP-violation in the dark sector, $C[t]$ is a model-dependent coefficient which we will take to be $1$ for simplicity,  
 and $\rho_{\rm DS}$ represents the energy density of a dark sector field in the current Universe. In Appendix.\,\ref{ap:1}, we show that various sources in concrete models of the dark sector can be represented in this form. We also show the case when $C[t]$ is varying in time.

To solve the equation from a general source, we first calculate the potential from a point source:
\beq
J^{\rm ps}(t,\vec{x})= 
\delta^3(\vec{x}).
\eeq
We obtain the solution to \Eq{eom} as
\beq
\laq{Force}
\d \f^{\rm ps}(t,\vec{x}) =\frac{1}{4\pi r}\exp{(-m_\f r)}  
\eeq
where $r=\ab{\vec{x}}.$
As a result, the force from a general non-relativistic, stationary, distribution can be given by the superposition of
\beq
\laq{potential}
\d \f(t,\vec{x})\approx \int{ d^3\vec{x}'\frac{\e \rho_{\rm DS}(x')}{f_\phi} \d\f^{\rm ps}(t,|\vec{x}-\vec{x'}|) }.
\eeq
The energy density of the dark sector, $\rho_{\rm DS}$, has ``charge" $\e/f_\f$ for the force. 
Note that even if $\e$ is non-vanishing, the force $\vec{\partial} \d\f$ would be vanishing if $\rho_{\rm DS}$ were spatially homogeneous. 
However, as we will see, the dark sector density is generally spatially inhomogeneous.

\subsection{Cosmic axion force acting on CP-preserving SM sector} 
We  can discuss the phenomena of the long-range force mediated by $\f$.
From the aforementioned reasons, we assume that the axion $\f$ has shift and CP symmetric interactions to SM fermions $\psi$. The lowest dimension term is given as  
\beq
\laq{int2}
{\cal L}= \frac{c_{\p}\partial_\mu \phi }{f_\f}\bar{\psi} \gamma_5\gamma^\mu \psi 
\eeq
where $c_{ \p}$ is a dimensionless constant. 
If there is such a term coupling to quarks, one obtains the couplings to the nucleons below the QCD scale
\beq
\to  \frac{c_{N}\partial_\mu\f}{f_\phi} \bar{N}\gamma_5 \gamma^\mu N.
\eeq
where $c_N$ is a constant which is related with $c_\p$ for quarks, 
 and $N=p, n$ is a nucleon. 
We emphasize that $\f$ is not a QCD axion. Since interaction \eq{int2} itself is completely shift symmetric, the QCD instanton does not generate the potential of $\f$. 
In other words, the shift symmetry is anomaly-free to the color gauge group~\cite{Pospelov:2008jk,Arias:2012az,Nakayama:2014cza,Takahashi:2020bpq, Bloch:2020uzh, Li:2020naa, Athron:2020maw, Han:2020dwo, Takahashi:2020uio}.

At the non-relativistic limit of the fermion, one obtains the interacting Hamiltonian as 
\beq
H\simeq - \frac{c_{i}}{f_\f} \vec{\partial} \d\f   \cdot \vec{\sigma}_i.
\eeq
Here $i= e, \mu, n, p$ etc. 
We can make a proper Lorentz transformation from/to this basis to get the hamiltonian in the relativistic limit.\footnote{Since $\d\f$ is time-independent, boosting a fermion does not enhance the spin precession frequency (it is enhanced in the rest frame by a Lorentz factor due to the Lorentz contraction, but it cancels out in the Laboratory frame due to the time duration). This is different from the fermion precession in an oscillating axion DM background. In this case, the spin precision is proportional to the velocity and thus the precession frequency is enhanced by a Lorentz boost \cite{Graham:2020kai}. 
On the other hand, a deuteron-like particle has a negative anomalous magnetic moment. 
In a storage-ring experiment of the particle, the spin can be frozen in the lab frame. It may be a good experimental tool searching for the cosmic axion force. }

The Hamiltonian resembles the one for magnetic moment interaction, $\mu_i \vec{B} \vec{\sigma}_i. $
One can identify the ``magnetic field" of \beq \vec{B}_{\rm eff}\equiv \frac{c_{ i} \vec{\partial} 
\d\f}{(\mu_i f_\f)}
\eeq coupled to the ``magnetic moment", $\mu_i \vec{\s}$. The neutron ($i=n$) has the value of $\mu_n\approx-1.9\mu_0$, a proton ($i=p$) has $\mu_p\approx 2.8\mu_0$, 
where $\mu_0\equiv e/2m_N \approx 0.1 ~e\cdot {\rm fm}$ is the nuclear magneton. For the charged lepton $i=e, \mu, \t$, $\mu_i\approx e/2m_i .$

\subsection{Cosmic axion force from DM} 
Let us give some concrete examples for the cosmic axion forces. A most important source perhaps is the DM.
Due to the primordial density perturbation and structure formation, the DM must be spatially inhomogeneous. 
Alternatively, we may also have compact objects formed by DM~\cite{Visinelli:2017ooc, Arvanitaki:2019rax} or topological defects such as domain walls~\cite{Kibble:1976sj, Kibble:1980mv, Takahashi:2020tqv}, which will be our later topic. 

We divide the energy density of the DM, $\rho_{\rm DS}$, into two parts, 
\beq
\rho_{\rm DS}=\rho_{\rm galactic}+\rho_{\rm extra}
\eeq
where $\rho_{\rm galactic}$ ($\rho_{\rm extra}$) is the energy density contribution from our galaxy by assuming some standard distributions (extra galactic component).

Let us adopt the NFW DM profile for $\rho_{\rm galactic}$~\cite{Navarro:1995iw,Cirelli:2010xx}. 
One can calculate the force contribution from 
\beq
\rho_{\rm galactic}=\rho_{\rm NFW}= \frac{\rho_s r_s}{r} \(1+\frac{r}{r_s}\)^{-2} 
\eeq
where $\rho_s\approx 0.184\GEV/{\rm cm}^3$, $r_s\approx 24.43\,$kpc, and the position of the sun is at  $r\approx r_{\odot}\approx 8.33$\,kpc. 
Then one can calculate the potential from \eq{potential} and obtain the cosmic axion force by taking the derivative. This contribution gives $B_{\rm eff}$ towards or opposes to the galactic center depending on the sign of $c_{i} \e $. This is because the DM distributes spherically around the galactic center in the NFW profile. 

In most of the space within the horizon, $\rho_{\rm extra}$ should be  
$\rho_{\rm crit} \times \Omega_{\rm DM}$ 
where the critical density of the Universe $\rho_{\rm crit}\approx 1.1 \times 10^{-5} h^2\GEV/{\rm cm}^3$ and $\Omega_{\rm DM}h^2\approx 0.12$~\cite{Aghanim:2018eyx}, with $h\approx 0.67$ being  the reduced Hubble parameter. An inhomogeneous distribution must exist due to the primordial density perturbation of $\O(10^{-3})\%$ from inflation~\cite{Aghanim:2018eyx}. 
We parametrize the inhomogeneity of $\rho_{\rm extra}$ as 
\beq
\vec{\xi}\equiv \(\rho_{\rm crit} \Omega_{\rm DM}  \frac{1}{3 m_\f} \)^{-1} \int{d^3\vec{x'}\rho_{\rm extra}(\vec{x'})\frac{\vec{x}-\vec{x'}}{|\vec{x}-\vec{x'}|^3}}  {\exp{(-m_\f |\vec{x}-\vec{x'}|)}}.
\eeq
If the dark sector distribution is spatially inhomogeneous $\xi\neq 0.$
We turn on this contribution if $r>r_s$. 
If $\rho_{\rm extra}$ contribution is dominant, we obtain an effective magnetic field of
$
\mu_{i}\vec{B}_{\rm eff} =\e \vec{\xi} \frac{c_i\rho_{\rm crit} \times \Omega_{\rm DM}}{f_\f^2 m_\phi}, 
$ 
which should increase if we decrease $ m_\f$ by fixing $f_\f$.
However, as we will see soon, this component is subdominant when $m_\f$ is greater than the Hubble constant.

\begin{figure}[t!]
\begin{center}  
\includegraphics[width=85mm]{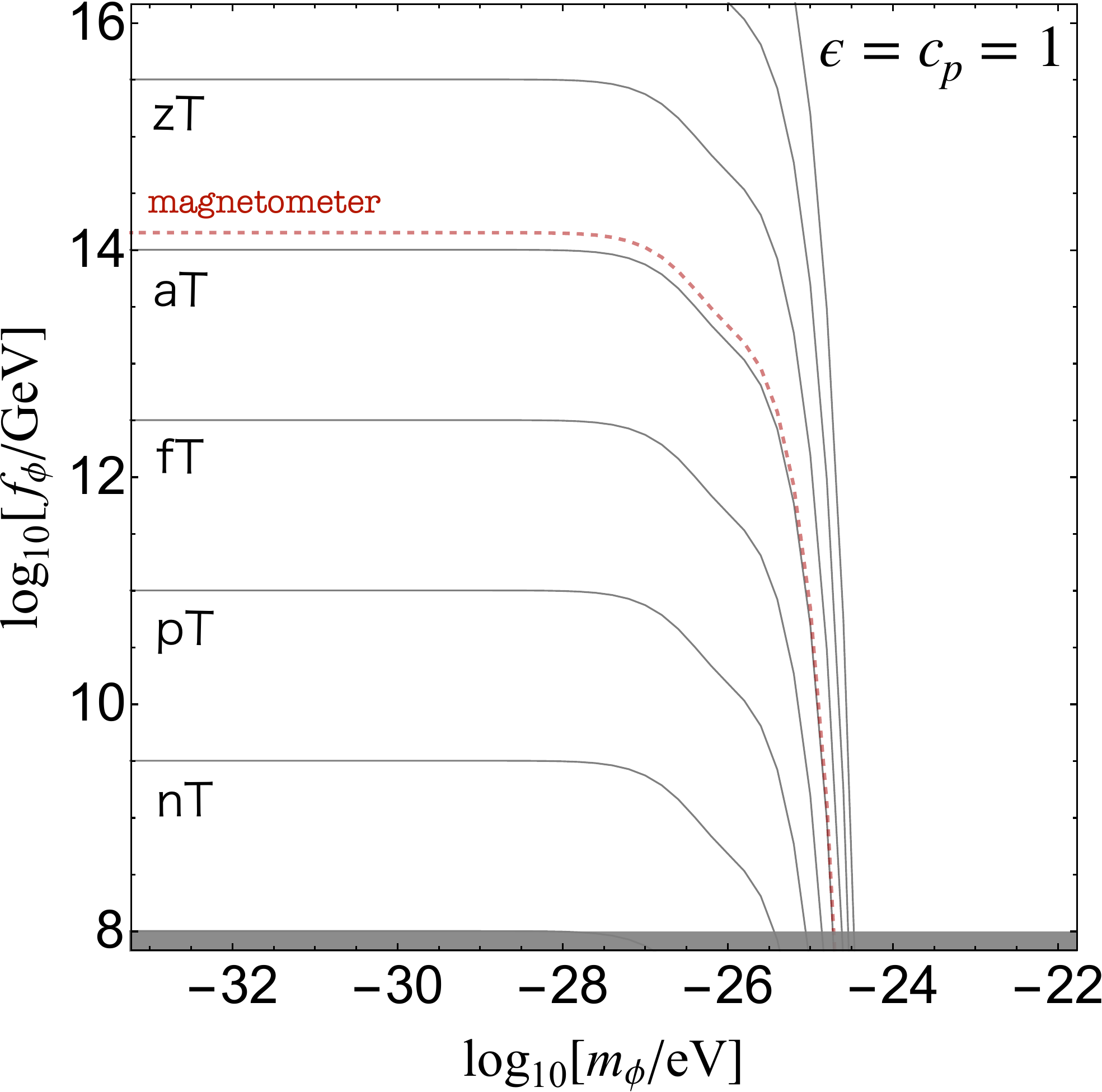}
\includegraphics[width=85mm]{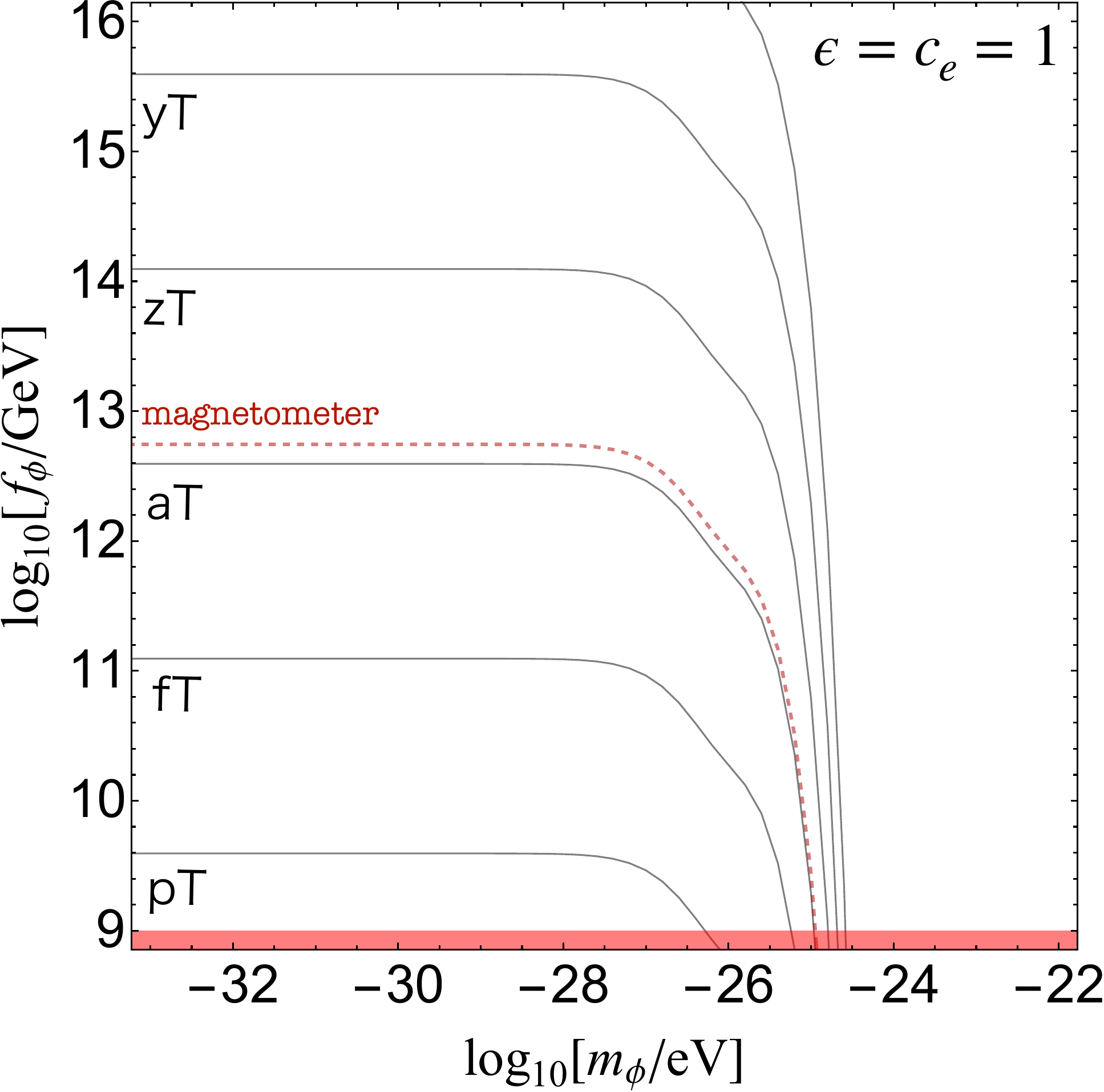}
\end{center}
\caption{  Contours of the effective magnetic field, $|\vec{B}_{\rm eff}|$ [T], for proton [upper panel] and electron [lower panel] from the cosmic axion force induced by the DM
 with $c_p=1$ and $c_e=1$, respectively. In both figures we take $\e=1$ and assume the NFW DM profile. 
 The direction is towards or opposes to the galactic center. 
 The lower colored region may be excluded by astrophysical bounds. 
 The red dotted curve represents $\simeq 0.5$aT, the sensitivity reaches for the magnetometers estimated in the next section.
 }
\label{fig:1}
\end{figure}

The cosmic axion force induces the spin precession of the SM fermions that couple to the axion. 
The effective magnetic field [$\rm T$] are given in Fig.\,\ref{fig:1} with $\x=10^{-5}$ for the nucleon and electron precessions in the upper and lower panels, respectively, with $c_p =1$ and $c_e=1$.
$\e=1$ is fixed.  
Red dotted contours represent the possible sensitivity reaches of magnetometers (See Sec.\,\ref{sec:3}). 
The shaded region in the bottom may be excluded by astrophysical bounds (see the following).
With general $\e, c_i$ the vertical axis can be regarded as $\log_{10}\(\sqrt{\(g^{(p)}_{a \psi_i \psi_i} g^{(s)}_{a{\rm DMDM}}\)^{-1} m_{\psi_i} m_{\rm DM}}/\GEV\)$ with $g^{(p)}_{a \psi_i \psi_i} =c_{\psi_i} m_{\psi_i}/f_\f$ and $g^{(s)}_{a{\rm DMDM}}=\e m_{\rm DM}/f_\phi$ (see also Appendix \ref{ap:1} for $g^{(s)}_{a{\rm  DMDM}}$). 
When $1/m_\f$ is much 
 larger than $r_{\odot}$, the induced magnetic field is almost constant. This is because the force from $r\gg r_{\odot}$ cancels out. 
This effect represents that the force is dominantly from the DM around the galactic center. 
When $1/m_\f$ is much larger than the size of our galaxy, on the other hand, the force from the density perturbation should become more important in principle. 
However, from the numerical estimation, we found that the extragalactic component is subdominant in the mass range shown in the figure. 
When $1/m_\phi$ is larger than the range shown in the figure, i.e. larger than the horizon size, the cosmic axion force from out of the Hubble horizon should not reach us.  
Therefore we can conclude that the robust prediction of this scenario is the direction of the effective magnetic field, which is towards or opposes to the galactic center.

If the dark sector particle, on the other hand, forms a compact object, the force direction might be biased to the object. 
We can use $\xi\gg 10^{-5}$ to estimate the size of the magnetic field for this effect. 
The formation of a compact object, the position and the typical $\xi$ are model-dependent. 
We will discuss the possibility of the cosmic axion force from an axion domain wall in Sec.\,\ref{sec:DW}.

\paragraph{Constraints on the DM scenario}
The axion is extremely light and it couples to the SM fermions. 
Such axion may be constrained from astrophysics and cosmology. 
Since the energy density or the potential height, $\rho_\phi \ll f_\phi^2 m_\phi^2$, is extremely small in the viable region ($f_\phi\lesssim 10^{14}\GEV$), 
 the constraints for the overproduction of $\f$ as DM or dark energy~\cite{Hlozek:2014lca} are negligible. 
However, there is an important constraint: the stellar cooling. 
If $\f$ couples to nucleon (electron), $ 
f_\f \gtrsim 10^8\GEV$ ($f_\f \gtrsim 10^9 \GEV$) is required from the cooling constraints of SN1987A~\cite{Mayle:1987as,Raffelt:1987yt,Turner:1987by, Chang:2018rso} (Red giant stars~\cite{Viaux:2013lha,Capozzi:2020cbu}). 
On the other hand, with $f_\f\lesssim 10^{8}\GEV$ one should also care axion monopole force induced by the nucleon in the QCD sector via CP-violation~\cite{Moody:1984ba, Pospelov:1997uv}.
If the strong CP phase, $\theta_{\rm CP}$, is non-vanishing, an axion-nucleon Yukawa coupling is induced
$g_{aNN} \sim \theta_{\rm CP} f_\pi/f_a $ with $f_{\pi}\sim 130\MEV$ being the  pion decay constant. 
In fact, the strong CP phase should not be completely vanishing and should satisfy $10^{-17}\text{-}10^{-16}\lesssim|\theta_{\rm CP}|\lesssim 10^{-10}$. Here the upper bound is from the non-observation of the neutron EDM and the lower bound comes from the electroweak contribution (by assuming a QCD axion)~\cite{Pospelov:2005pr}. 
Even if we introduce the QCD axion to solve the strong CP problem, a CP-violating week interaction drives the minimum of the QCD axion slightly away from the CP-conserving place. 
By taking into account the CP-violating effect, 
the nucleon also induces a monopole potential, which is constrained by the aforementioned test of monopole-monopole interaction~\cite{Schlamminger:2007ht} and monopole-dipole interaction~\cite{Venema:1992zz} between visible particles (See also \cite{Adelberger:2009zz,Jaeckel:2010ni, Safronova:2017xyt}).
We get the constraint $ f_\f\gtrsim 10^7\GEV~ {\rm if } ~\theta_{\rm CP}=10^{-17}$ for $m_\f\lesssim 10^{-15}\EV.$
We find that the reach of $f_\f\sim 10^{14}\GEV$ for our monopole-dipole interaction between dark and visible sectors can be many orders of magnitudes beyond those set by the astrophysical and ordinary long-range force constraints. 

The reason that the cosmic axion force from DM can be very strong is that the force is from a constructive superposition over a cosmological scale. This is similar to the fact that we feel gravity despite it is extremely weaker than the electromagnetic force. The reason that the interaction between dark and visible sectors is less constrained than the one between visible sectors is the difficulty to measure the dark sector interaction. Indeed, the constraint on DM monopole-monopole interaction is even weaker~\cite{Ackerman:mha}. For instance, the asymmetric DM from dark baryon~(See the Appendix \ref{ap:1}), who may have mass around $m_{N'}=\O(1)\GEV$, should satisfy $f_\f\gtrsim 10^{-2} m_{N'}.$

\subsection{Cosmic axion force from domain wall} 
\label{sec:DW}
A simple example for a compact object or a topological defect is a domain wall. 
There are various mechanisms to form the axion domain wall~\cite{Kibble:1976sj, Kibble:1980mv, Takahashi:2020tqv}. 
The interesting point for our scenario is that the interaction between an axion and a axion domain wall is generically CP-violating (see the Appendix \ref{ap:1}).
If the axion domain wall follows the so-called scaling solution, we expect $\O(1)$ domain walls within a Hubble horizon. 
To evade the domain wall problem, the tension of the domain wall should satisfy~\cite{Zeldovich:1974uw,Vilenkin:1984ib} 
\beq
\sigma_{\rm DW} \lesssim (1\MEV)^3.
\eeq

If the axion domain wall couples to another axion, $\f$, the long-range force is emitted from the wall.  
The limited amount of the domain walls implies that the force is difficult to be canceled out unless we tune the position and direction of several domain walls. 
In the following we consider for simplicity that there is only a single domain wall inside our Hubble horizon and stretch over perpendicular to the position vector, $\vec{r}_{\rm DW}$. 

The potential from a non-relativistic domain wall can be calculated as 
\beq
\laq{potentialDW}
\d \f(t,\vec{x})\approx \int{ d^2\vec{x}_\perp\frac{\e \s_{\rm DW}}{f_\phi} \d\f^{\rm ps}(t,|\vec{x}-\vec{x'}|) }.
\eeq
where the integral is performed on the surface of the domain wall. 
Then we obtain, analytically, 
\beq
\d \f(t,\vec{x})= \e \frac{\sigma_{\rm DW}}{f_\phi} \frac{e^{-m_\f |\vec{r}_{\rm DW}|}}{2 m_\f}.
\eeq
The force is perpendicular to the wall and behaves as an effective magnetic field, e.g. for a proton as
\beq
|\vec{B}_{\rm eff}|= 3{\rm \, aT} \times {\g_{\rm DW}} \e  c_p e^{-|\vec{r}_{\rm DW}| m_\f}  \(\frac{10^{12}\GEV}{f_\f}\)^2 \frac{\s_{\rm DW}}{\(1\MEV\)^3},
\eeq
where $\gamma_{\rm DW}\equiv 1/\sqrt{1-(\dot{\vec{r}}_{\rm DW})^2}$ is the Lorentz factor for the domain wall motion, which takes into account the Lorentz contraction and may enhance the magnetic field. 
This is exponentially suppressed if $|\vec{r}_{\rm DW}|$ is larger than  $1/m_\f.$ For the cosmic axion force mediated by a lighter axion than $1/|\vec{r}_{\rm DW}|$, the effective magnetic field does not 
depend much on the distance to the domain wall. Therefore, the cosmic axion force can be mediated to Earth from an extremely distant domain wall. 
 
We may wonder if the axion, $a$, forming the domain wall can also mediate the force.  
Indeed, there is a force mediated by $a$. That is nothing but the gradient of domain wall configuration. The corresponding magnetic field is now being searched for in GNOME experiment~\cite{Pustelny:2013rza,Afach:2018eze,Afach:2021pfd}. 
The experimental advantage of introducing another light axion is that the force range, $1/m_\f$, can be much longer than the wall-size $1/m_a$ with $m_\phi \ll m_a$. 
Then the probability for the interaction with the detector on Earth is extremely enhanced. 
The disadvantage may be that the magnetic field is almost a constant value unless $|\vec{r}_{\rm DW}|\sim 1/m_\f$ and domain wall is moving.

\section{Measurement of cosmic axion force}
\label{sec:3}
So far we have shown that the cosmic axion force can theoretically exist, behaves as the effective magnetic field, 
which induces the spin precession of the SM fermions. 
The precession frequency is given as, e.g.,
\beq
\laq{freele}
f_{\rm electron} =  2.4 \text{day}^{-1}\times \frac{|\vec{B}_{\rm eff|}}{\rm fT}
\eeq
\beq
f_{\rm proton} =  1.3 \text{year}^{-1}\times \frac{|\vec{B}_{\rm eff}|}{\rm fT}
\eeq
for electron and proton, respectively.

This magnetic field features the following properties:
\begin{itemize}
\item The effective magnetic field is kept intact even if we shield all the ordinary magnetic fields away (See the discussions in  Sec. \ref{sec:shield} for electron coupling).
\item The effective magnetic field is towards a fixed direction, as the galactic center, everywhere on Earth. 
\item  The effective magnetic field is almost constant within an experimental timescale say $>\O(1)$~yr.
\end{itemize}
By keeping in mind those properties we discuss the testability of the cosmic axion force in this section.

\subsection{Magnetometers and daily modulation}

A magnetically shielded atomic magnetometer could be a detecter for the pseudo-magnetic field, $\vec{B}_{\rm eff}.$
It effectively screens conventional magnetic fields for a whole day, and the axionic monopole signals can be accumulated. However, usually, an atomic magnetometer has directional sensitivity and this direction will have a precession unless it is orthogonal to the axial tilt direction. Here we consider the detector directing along the celestial equator, for simplicity, to reduce precession error.

When the magnetometer is operating, the axionic monopole gives a constant signal for a short period. 
By taking into account the Earth's rotation, the signals have daily modulation. 
 Figure~\ref{fig:scheme} illustrates a conceptual diagram of the possible monopole (black) search using the atomic magnetometer as a detector (red) in the Earth (blue).  The angle between the monopole position vector from the detector and its sensitive direction is $\theta$. Thus the daily modulation depends on $\theta$. 
Roughly speaking, the signal amplitude is reduced by a factor of $\cos\left(\theta\right)$. Then the daily modulation peak-to-peak amplitude will be
\begin{equation}
\delta B\approx 2\left|\vec B_\mathrm{eff}\right|\cos\theta,
\end{equation}
where the cosine term gives
$
\cos\theta\geq0.1\quad\mathrm{for}\quad\theta\leq84.3^\circ.
$
For practical case, one has
\begin{equation}
\delta B\geq 0.2\left|\vec B_\mathrm{eff}\right|.
\end{equation}

 In the following we estimate the sensitivity, by taking care of the daily modulation by taking $\cos \theta=\O(1).$ 
We note that since we know the direction of the galactic center (the galactic longitude and latitude $\sim 0$), we can optimize the direction of the detector when we measure the force from DM.

\subsection{Sensitivity estimation}
Depending on the experimental design, the averaging method can be used in two ways; incoherent averaging, and coherent averaging (See Appendix. \ref{app:2}). The dependency of experiment repetition number $N$ for a signal to noise ratio (SNR) is different depending on the averaging type as:
\begin{equation}
\begin{split}
&\mathrm{SNR_{p}^{i}}=\left(\frac{B_{\mathrm{sig}}}{B_{\rm noise}}\right)^{2}\sqrt{N}\\
&\mathrm{SNR_{p}^{c}}=\left(\frac{B_{\mathrm{sig}}}{B_{\rm noise}}\right)^{2} N,
\end{split}
\end{equation}
where the superscripts $i,\ c$ denote the incoherent averaging and the coherent averaging respectively. Here we used the SNR as power, which is square of the field SNR. For white noise, the $B_{\rm noise}$ can be written with power spectral density ($\delta B_{n}$) with unit of $\rm{T}/\sqrt{Hz}$ as:
\begin{equation}
B_{\rm noise}=\delta B_{n}\sqrt{2b},
\end{equation}
where $b$ is resolution bandwidth, and numerical factor 2 comes from the Nyquest theorem. For example, the nucleon-electron co-magnetometer has sensitivity order of $1\,\rm{fT}/\sqrt{Hz}$ range in the low frequency region less than $1\,\mu\rm{Hz}$~\cite{PhysRevLett.89.253002, PhysRevLett.95.230801, PhysRevLett.105.151604}. Since the cosmic axion force has a daily modulation,  one can use the coherent averaging method. For target ${\rm SNR}=7$, repetition number 300, and resolution bandwidth $10\,\mu\rm{Hz}$ for 1 day time series data as a single measurement, the detectable signal strength is,
\begin{equation}
B_{\mathrm{sig}}\approx 0.5\,\mathrm{aT}\left[\frac{7}{\mathrm{SNR_{p}^{c}}}\right]^{0.5}\left[\frac{N}{300}\right]^{0.5}\left[\frac{1\,\mathrm{fT/\sqrt{Hz}}}{\delta B_{n}}\right]\left[\frac{10\,\mu\mathrm{Hz}}{b}\right].
\end{equation}

\subsection{On magnetic shielding and force direction measurement}
\label{sec:shield}
Most atomic magnetometers use either alkali metals or noble gases. Those magnetometers have two steps (if one omits optical pumping); first, spin precession motion with angular frequency $\omega$ under magnetic field $B$
\begin{equation}
\omega=\gamma B,
\end{equation}
where $\gamma$ is the gyromagnetic ratio of an optical medium. The second step is probing a polarization by dichroism or birefringence.

We have started with an atomic magnetometer having an exotic electron spin coupling to a light axion. However, since the axion force also acts on the electron in a ferro-/ferri-magnetic shielding,
this electron coupling could be screened~\cite{PhysRevD.94.082005}. Since they are conventionally used in atomic magnetometers, the effective magnetic field seen by an electron will be suppressed by $\O(100)$ if this effect fully occurs. 
{However, we argue that this problem may be avoided if the effective magnetic field is in sub fT range.
The time-scale of the induced motion of an electron spin is longer than a day (see \Eq{freele}), i.e. the induced magnetic field for the shielding is in the time-averaged force direction in the Laboratory frame. 
However, the (daily modulating) spin precession may be around a different direction than the original force direction due to the shielding. 
In this case, we may measure the force direction with several detectors at different places like in GNOME. 
Alternative simple possibility to avoid this problem is to rotate the shield much faster than $f_{\rm electron}$ while keeping the magnetometer intact.}

Instead, nucleon coupling is not affected by the magnetic shielding and we do not need to care this issue. 
From the daily modulation of the nucleon spin motion, we can measure the force direction. 
 It will be a smoking-gun evidence of our scenario if the force direction is towards the galactic center.

\section{Conclusions and discussion}

We have studied a very light axion as a mediator of a long-range force between the dark sector and visible sector. 
The key assumption is that the dark sector, from which the force originates, has CP-violation. 
Even if the dark sector component is extremely far away, it can affect the Laboratory detector on Earth via the long-range force. The force behaves as a magnetic field which induces the spin precessions of nucleons or leptons. 
The precession is around a fixed direction and is kept even with an ordinary magnetic field shielding. 
Such precessions can be detected in magnetometers via the daily modulation of signals. 
The constant effective magnetic field towards the galactic center is a smoking gun prediction of the cosmic axion force from the dark matter. 

Let us mention a few extensions of our proposal.
First of all, a similar scenario can be obtained if the dark sector is charged under a dark abelian gauge group, whose gauge field couples to the visible particles via electric dipole moment operators. 
Another interesting possibility is that the axion $\f$  forms non-trivial distribution around an ordinary star like in Ref.\,\cite{Hook:2017psm}, which discusses  finite density corrections for the QCD axion potential in a neutron star. 
In our case, $\f$ does not get the potential from QCD instanton and this effect is neglected. 
However, the Peccei-Quinn (PQ) symmetry can be restored inside a star 
if the PQ field for the axion is light enough compared to the finite density effect inside a star. 
The PQ field and matter coupling may be caused by the mixing between the PQ field and the Higgs field.  
Then, the axion field may obtain a non-trivial distribution around the star, and thus the gradient of $\f$ behaves as the cosmic axion force. 
In this case the force direction may be almost towards or opposes to Sun, especially when $1/m_\f$ is not too larger than the distance between Earth and Sun.

\section*{Acknowledgments}
W.~Y would like to thank the members at Center for Axion and
Precision Physics Research for kind hospitality when this work was initiated and for the warm help when he was in difficulty due to COVID. 
 D.~K, Y.~K, Y.C.~S,  and Y.K.~S were supported by the IBS-R017-D1-2021-a00 of the Republic of Korea.
 W.~Y was supported by JSPS KAKENHI Grant Nos. 16H06490, 19H05810 and 20H05851.

\appendix 
\section{Dark sector models with CP-violating axion coupling }
\label{ap:1}

Let us explain that the axion coupling  to the dark sector  may easily violate CP. 
To restrict ourselves, let us consider that  the theory has CP symmetry at high energy scales (at the perturbative level). 

\subsection{Dark QCD with a strong CP phase}
For instance, we can consider a QCD-like model (we call it dark QCD) with a non-vanishing strong CP phase coupled to the axion with \eq{shift}, 
\beq
{\cal L}\supset - \theta'_{\rm CP} \tl{G'} G'- \frac{c_{\p'}\partial_\mu \phi }{f_\f}( \bar{u'} \gamma_5\gamma^\mu u'+ \bar{d'} \gamma_5\gamma^\mu d' ) 
\eeq
where we have neglected to write down the kinetic (and mass) terms.  Again we notice that the axion does not solve the strong CP problem for this model since the Lagrangian is shift symmetric under $\f \to \f+ \a$ and $\f$ never has a potential to eliminate  $\theta'_{\rm CP}.$
Here we have assumed that the dark quarks have two flavors,  and axion couples to them universally. 
This is quite similar to the ordinary two-flavor QCD except for the CP phase and the axion couplings. 
Thus we expect confinement for large enough gauge coupling. 
Then dark nucleon, $N'$, will have a CP-violating Yukawa interaction, ${\cal L}_{\rm eff} \supset -g_{\f N'N'} \f \bar{N'} N'$, of the form with~\cite{Moody:1984ba, Pospelov:1997uv}
\beq 
g_{ \f N'N' }= \frac{\theta_{\rm CP}' c_{\psi'} }{f_\f} \frac{ m_{u'}m_{d'}}{ \(m_{u'}+m_{d'}\)} \langle N'| \bar{u}'u'+\bar{d}'d' |N'\rangle.
\eeq
If  $m_{u'} \sim m_{d'} \sim \langle N'| \bar{u}'u'+\bar{d}'d' |N'\rangle  \sim m_{N'}$, with $m_{N'}$ being the mass of the dark nucleon,  we obtain the CP-violating coupling of order 
\beq
g_{ \f N'N' }\sim c_{\p'} \theta_{\rm CP}' \frac{m_{N'}}{f_\f}.
\eeq
For comparison, we mention that in the ordinary QCD, due to the fine-tuning of the strong CP phase, this term is small. However, if $\theta'_{\rm CP}$ is not finely tuned, we have  CP-violating coupling between axion and the nucleon.
 The dark nucleon may compose the asymmetric DM, 
$\e \sim \theta_{\rm CP}' c_{\psi'} $ in this case, and $\e=\O(1)$ if there is no tuning. 
From the asymmetric DM abundance, $m_{N'}=\O(1)\GEV$.

\subsection{Spontaneous CP breaking in axiverse}
In axiverse, many axions, $\f_i$, have a potential generated by a non-perturbative effect:
\beq
V=V(\f_i/f_i),
\eeq
where $f_i$ is the decay constant. 
The axions enjoy discrete shift symmetry \beq 
\f_i/f_i\rightarrow \f_i/f_i+ 2\pi,
\eeq
under which the potential is invariant, 
\beq
V(\f_i/f_i)=V(\f_i/f_i+2\pi).
\eeq
The periodicity implies that the potential can be given in the form
\beq
V=-\L^4 \sum_{n_j} \k_j \cos{\(\sum_{i}n^j_i {\f_i \o f_i} +\theta_j\)},
\eeq
where $n_i^j$ are integers. (We omit the constant term to cancel the vacuum energy today.) Immediately, we find that there are CP-phases,  $\theta_j$. 
Therefore there could be CP violation in general.

Even if we take the ``CP-symmetric" limit, $\theta_j=0$, 
the CP symmetry can be spontaneously broken. For simplicity, we consider a two axion model with potential given by
\beq 
\laq{orig}
V= -\L^4\(\k_1\cos(n_a\frac{a}{f_a}) +\k_2\cos(\frac{\phi}{f_\f}+\frac{a}{f_a}) +\k_3\cos(n_\f\frac{\phi}{f_\f})\)\eeq
 with $\k_1\gtrsim \k_2,\k_3, n_a>1$ then we can integrate out $a$ who has local minima $\vev{a}/f_a\approx 0,2\pi/n_a,4\pi/n_a\cdots (n_a-1)2\pi/n_a.$ 
 The mass of $a$ is 
 \beq
 m_a^2\sim \k_1 \frac{n_a^2\L^4}{f_a^2}
 \eeq
Except for the first minimum, we obtain a non-vanishing CP phase from the 
spontaneous symmetry breaking. 
This appears 
in the low energy theory as \beq V_{\rm eff}\simeq  -\L^4 \(\k_2\cos(\theta+\frac{\f}{f_\f})+\k_3\cos(n\frac{\f}{f_\f})\)\eeq
where $\theta\equiv\vev{a}/f_a.$\footnote{This kind of potential is known to lead to a consistent inflation~\cite{Czerny:2014wza,Czerny:2014qqa,Croon:2014dma,Higaki:2015kta,Daido:2017wwb,Daido:2017tbr,Takahashi:2019qmh,Takahashi:2020uio}. }
Let us expand the potential around the minimum of $\f$. 
We obtain
\beq
V_{\rm eff}\simeq {m_\phi^2 \o 2}  \d\f^2+ {A_\f\o 3!} \d\f^3+\cdots
\eeq
Here $\d\f\equiv \f-\vev{\f}$, with $\vev{\f }/f_\f\simeq  -\theta \k_2/(\k_2+n^2\k_3)$, \beq m_\phi^2\simeq  (\k_2+n^2\k_3) \frac{\L^4}{f_\f^2} ,\eeq and \beq
A_\f \simeq   \theta\frac{ \k_2 \k_3 n^2}{\k_2+\k_3 n^2}\left(1-n^2\right)\frac{\L^4}{f_\f^3}+\O(\theta^3)\equiv \e_\f \frac{m_\f^2}{f_\f},
\eeq
which is non-vanishing if $\theta\neq 0.$ This term is obviously CP-violating. 
This will give a source term for the long-range force discussed in the next subsection. 

Let us come back to the original potential \Eq{orig}, by defining $\d a\equiv a-\vev{a},$ we obtain the interacting term from the second cosine term as 
\beq
V\supset \frac{A_a}{2} \d a^2 \d \f
\eeq
where 
\beq
A_a\simeq \theta \frac{\k_2 \k_3 n^2 }{\k_2+\k_3 n^2}\frac{ \L^4}{f_a^2 f_\f} +\O(\theta^3)\equiv \e_a \frac{m_a^2}{f_\f}.
\eeq
Consequently, other than the self-cubic-interaction, the CP-breaking also induces the cubic interaction between heavier, $a$, and lighter, $\f$, axions. 
We mention that $a$ can be the QCD axion if it is anomalously coupled to the gluons. In this case we may identify $\k_1 \L^4$ as the topological susceptibility $\k_1 \L^4/n_a^2\sim (0.0756\GEV)^4$~\cite{Borsanyi:2016ksw}. 
The quality problem can be solved if $\k_2$ is small enough~\cite{Acharya:2010zx} or induce testable EDM in the proton EDM experiment if  
$\k_2\L^4 \sim (0.3\MEV)^4$~\cite{Marsh:2019bjr}. For large enough $\k_3n^2$ we obtain $\epsilon_a \sim \O(10^{-10})\theta$, which induces the magnetic field of $\rm \O(1) aT$ for $f_\f=10^{8-9}\GEV, \theta=\O(1) \AND c_p=1$.
If $a$ and $\f$ are both non-QCD axions with $\k_1\sim \k_2 \sim \k_3$, $m_a\gg m_\f$ can be made by $f_a\ll f_\phi$. In this case, $\e_a\sim \O(\theta ), $ which is order $1$ for $n_a=\O(1).$

 If $\f$ or $a$ contributes to the density of the Universe (either as DM does or whatever else), 
 as long as they are non-relativistic, one can approximate
\beq \{\d\f,\d a\}(t,\vec{x})= \sqrt{\frac{\rho(\vec{x})}{m^2}}\cos[m t],\eeq
where $\rho(\vec{x})=\{\rho_\phi(\vec{x}),\rho_a(\vec{x}) \}$ ($m=\{m_\phi, m_a\}$) represents the density (mass) of $\{\f, a\}$.
We would like to obtain the axion potential from source $J$. 
 Here we obtain
\beq
J= \{A_\f \d\f^2, A_a \d a^2\}
\eeq
Then we can derive the solution to \eq{eom} as 
\beq
\laq{Force}
\d \f^{\rm ps}(\vec{x},t) =  \(\frac{1}{4\pi r}\exp{(-m_\f r)}+ \frac{\cos{(r\sqrt{4m^2-m_\f^2} )}\cos{(2m t)}}{4\pi r}\).
\eeq
from a point source of the form 
\beq
J^{\rm ps}(t,\vec{x})=\(\cos{[m t]}\)^2\delta^3(\vec{x})=(1+\cos[2mt]) \delta^3(\vec{x}).
\eeq 
The oscillation term (2nd term) is not important if we consider $r\gg 1/m.$

\subsection{CP-violating interaction with axion domain wall}
Even if the vacuum is CP-conserving, we may still have a topological defect that carries the charge of the long-range force. 
To see this,  let us take $n_a=n_\f=1,$ in which case  the vacuum is $\vev{\phi}=\vev{a}=0,$ i.e. CP-conserving. 
Consider an $a$ domain wall configuration,  $a_{\rm DW}$, which satisfies $a_{\rm DW}[x,y,z\approx z_{\rm DW}]/f_a\sim \pi \mod 2\pi$ with $z_{\rm DW}$ being a position for a domain wall stretching in $x-y$ plane. 
At other $z$, $a_{\rm DW}/f_a$ takes vacuum value  $\vev{a}/f_a \sim 0 \mod 2\pi.$

 We can obtain
 \beq
 \frac{\partial{V}}{\partial\d\phi}\approx  -\k_2 \L^4 /f_\f \sin[a_{\rm DW}/f_a]+\O(\d \f/f_\f)
 \eeq
 The domain wall width is around $ 1/m_a,$ and thus only within the region $|z- z_{\rm DW}|\lesssim 1/m_a$, the r.h.s. $\sim \pm \k_2\L^4 /f_\f$, otherwise zero. 
  If $1/m_\f\gtrsim 1/m_a$ we can approximate 
\beq
\ab{ \frac{\partial{V}}{\partial\d\phi}}\sim \delta{(z-z_{\rm DW})} \frac{ |\k_2|\L^4}{f_\f} \frac{1}{m_a}\sim \delta{(z-z_{\rm DW})} \frac{\sigma_{\rm DW} }{f_\f}\ab{\frac{\k_2}{\k_1}}.
\eeq
Here the tension of the domain wall  is given by $\s_{\rm DW}\sim f_a^2 m_a.$
Since $\rho_{\rm DS}=\sigma_{\rm DW}\delta(z-z_{\rm DW})$, $|\e|\sim |\k_2/\k_1|.$

\section{Incoherent and coherent averaging methods}
\label{app:2}
\subsection{Number of repetition dependency on signal-to-noise ratio (SNR)}
\begin{figure}[h]
\centering\includegraphics[width=0.8\textwidth]{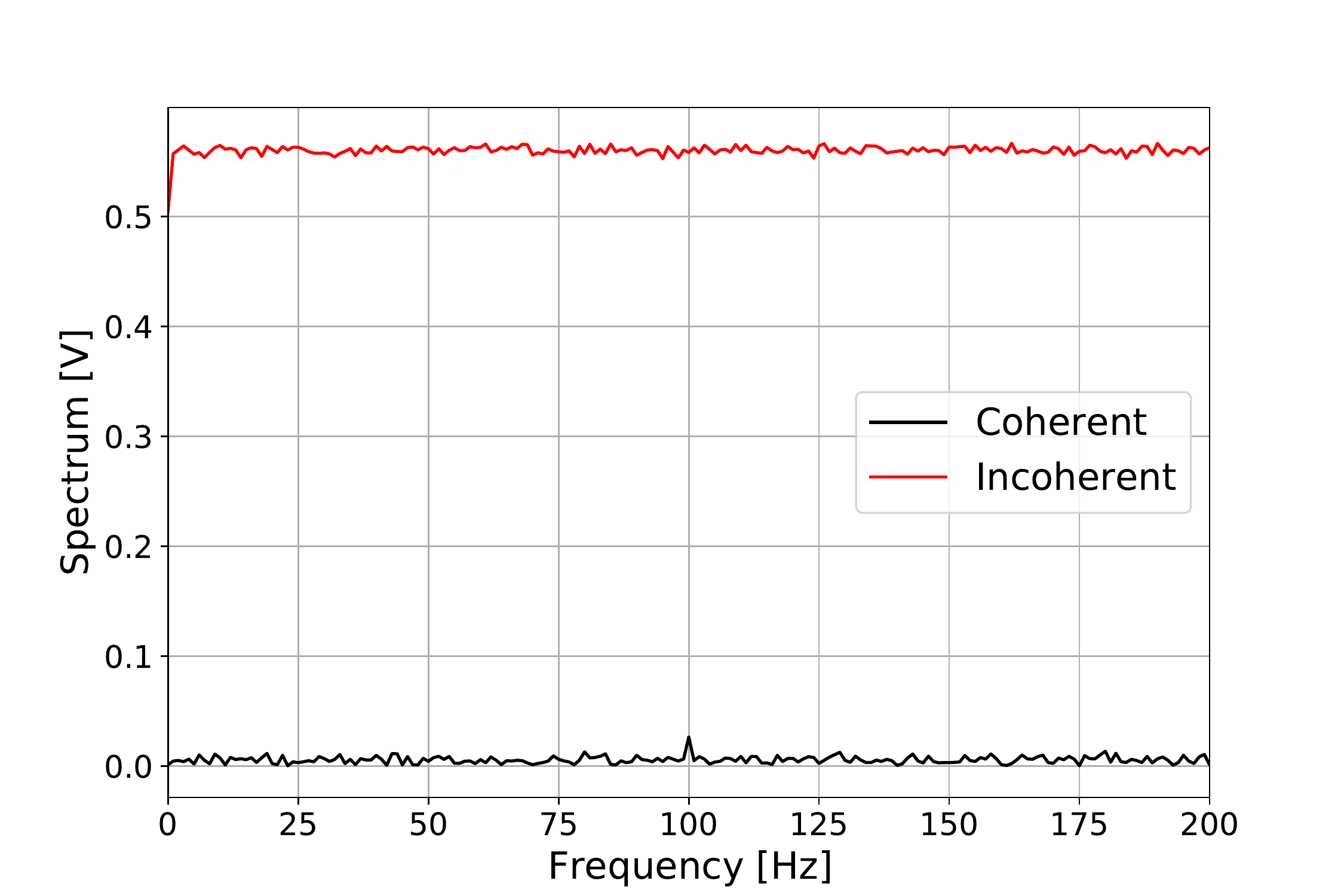}
\caption{Simulation of the spectrum depending on the averaging method.}
\label{fig spectrum_sim}
\end{figure}

\begin{figure}[h]
\centering
\begin{subfigure}[h]{0.49\textwidth}
\includegraphics[width=\linewidth]{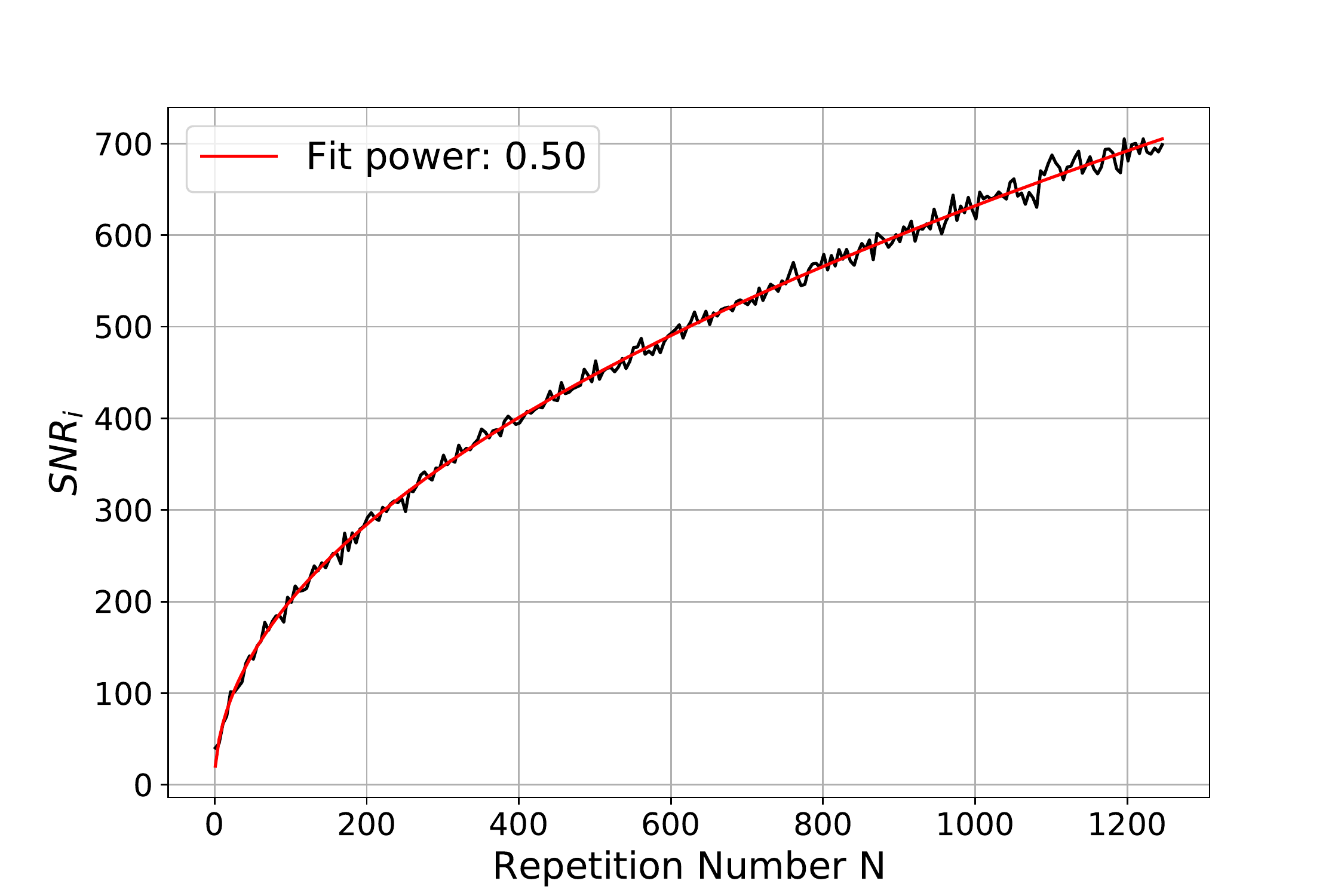}
\caption{}
\end{subfigure}
\begin{subfigure}[h]{0.49\textwidth}
\includegraphics[width=\linewidth]{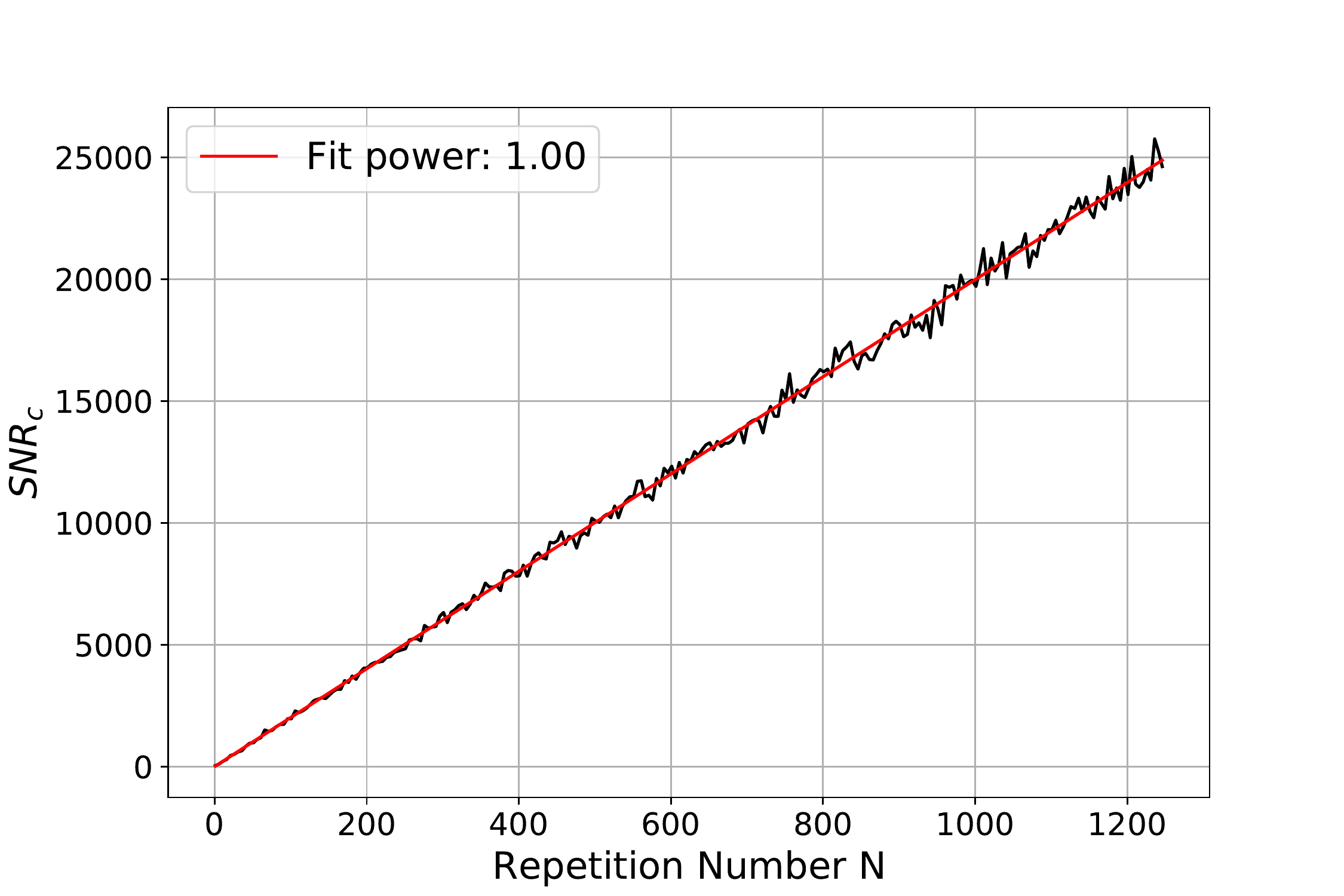}
\caption{}
\end{subfigure}
\caption{(a) SNR dependency to repetition number $N$ for incoherent averaging (b) SNR dependency to repetition number $N$ for coherent averaging}
\label{fig simulate SNR/N}
\end{figure}

The dependency of repetition number for SNR can be estimated by using two different averaging methods. 
For instance, we generate the sinusoidal signal with a frequency $100\,$Hz and amplitude $0.02\,\rm{V_{rms}}$, and $0.2\,\rm{V^{2}/Hz}$ of white noise with a sampling rate 10\,kHz. This single test set is denoted as $x_{i}$. This test set is generated $N$ times. Therefore there are total $N$ sets of the times series test samples, $X=\{x_{1},x_{2},\cdots,x_{N}\}$. Depending on the averaging method, the Fourier series coefficient is different as follows:
\begin{equation}
\begin{split}
&\tilde{X}_{c}=\mathcal{F}\left[\langle X\rangle\right],\\
&\tilde{X}_{i}=\langle\mathcal{F}\left[X\right]\rangle,
\end{split}
\label{eq coherent incoherent relation1}
\end{equation}
where $\mathcal{F}$ is the Fourier transform operator acting on the time series data, and $\vev{Y}$ represents the expectation value of $Y$. Since the former method averages the time series data first, it makes to preserve the phase information of the signal. However, the latter average method transforms the time series data to the Fourier space, and it makes to lose the phase information. For the same time series data $X$, $\tilde{X}_{c}$, and $\tilde{X}_{i}$ can be calculated. Fig.~\ref{fig spectrum_sim} shows the two different spectrum.
In this calculation, the repetition number is set to $N=10^{4}$. We see that the coherent averaging method gives the clear signal spectrum, but the incoherent averaging method does not. Since the coherent averaging reduces the noise energy spectrum, therefore it makes the overall noise level drop. On the other hand, the incoherent averaging maintains the noise energy spectrum, and the averaging process only reduces the fluctuation at the same noise level. Therefore the dependency of SNR for repetition number can be calculated with enough high injection signal strength($2\,\rm{V_{rms}}$ amplitude sinusoidal wave with 100\,Hz). Figure~\ref{fig simulate SNR/N} shows the power SNR dependency to $N$.
The calculated SNR(black) is fitted using a model function $f(N)=aN^{b}$, and the fitting parameter $b$ is labeled in Fig.~\ref{fig simulate SNR/N}. The power SNR of incoherent averaging grows with $\sqrt{N}$, and that of coherent averaging increases with $N$, which showed the expected behavior.

\subsection{Expected noise probability distribution}
A study for  the distribution of white noise spectrum is conducted by a simulation of coherent averaging with $N$ repetitions. Figure~\ref{fig spectrum_sim} shows that the noise spectrum fluctuates along the suppressed noise level near zero. The expected probability distribution for the power spectrum of coherent averaging and that of incoherent averaging can be estimated. In this estimation, the noise spectrum is white. Then the single time series noise $x_{i}\in X$ follows the normal distribution $N(0,\sigma)$.

First, the probability distribution of incoherent averaging will be estimated as follows. From Eq.~(\ref{eq coherent incoherent relation1}), incoherent averaging Fourier transforms each time series and then averaging the Fourier component. The final distribution can be traced by tracking individual steps. The Fourier transform of $N(0,\sigma)$ distribution can be separated into real part and imaginary part. Each component follows the $N(0,\sigma\sqrt{2/L})$, where $L$ is the length of the $x_{i}$. Therefore the power spectrum of $x_{i}$ will follow the probability distribution $\mathcal{T}_{i}\sim N(0,\sigma\sqrt{2/L})^{2}+N(0,\sigma\sqrt{2/L})^{2}$. The individual distribution $\mathcal{T}_{i}$ can be reduced to $\mathcal{P}_{i}=\mathcal{T}_{i}/\left(\sigma\sqrt{2/L}\right)^{2}$ which is the chi-square distribution with a degree of freedom of 2. Then we average this distribution with repetition number $N$. In symbolic notation,
\begin{equation}
\mathcal{T}=\frac{1}{N}\sum_{i}^{N}\mathcal{T}_{i}=\frac{1}{N}\sum_{i=1}^{2N}N(0,\sigma\sqrt{\frac{2}{L}})^{2}=\frac{2\sigma}{LN}\sum_{i=1}^{2N}N(0,\sigma)=\frac{2\sigma}{LN}\chi^{2}_{2N}.
\end{equation}
Therefore if we normalize the power spectrum to $2\sigma/(LN)$,  it follows the chi-square distribution with a degree of freedom $2N$.

The probability distribution for coherent averaging, on the other hand, can be estimated similarly. Coherent averaging conducts the expectation operation to the time series. Therefore the averaged time series $\langle x\rangle$ follows the distribution $N(0,\sigma/\sqrt{N})$. The Fourier transform operation to $\langle x\rangle$ makes power spectrum, and this spectrum will follow the distribution $\mathcal{T}_{c}$ as
\begin{equation}
\mathcal{T}_{c}=N(0,\sigma\sqrt{\frac{2}{NL}})^{2}+N(0,\sigma\sqrt{\frac{2}{NL}})^{2}=\frac{2\sigma}{LN}\chi^{2}_{2}.
\end{equation} 
This implies that a normalized power spectrum of the coherent averaging method follows the chi-square distribution with degree of freedom 2.

\begin{figure}[h]
\begin{subfigure}[h]{0.49\textwidth}
\includegraphics[width=\linewidth]{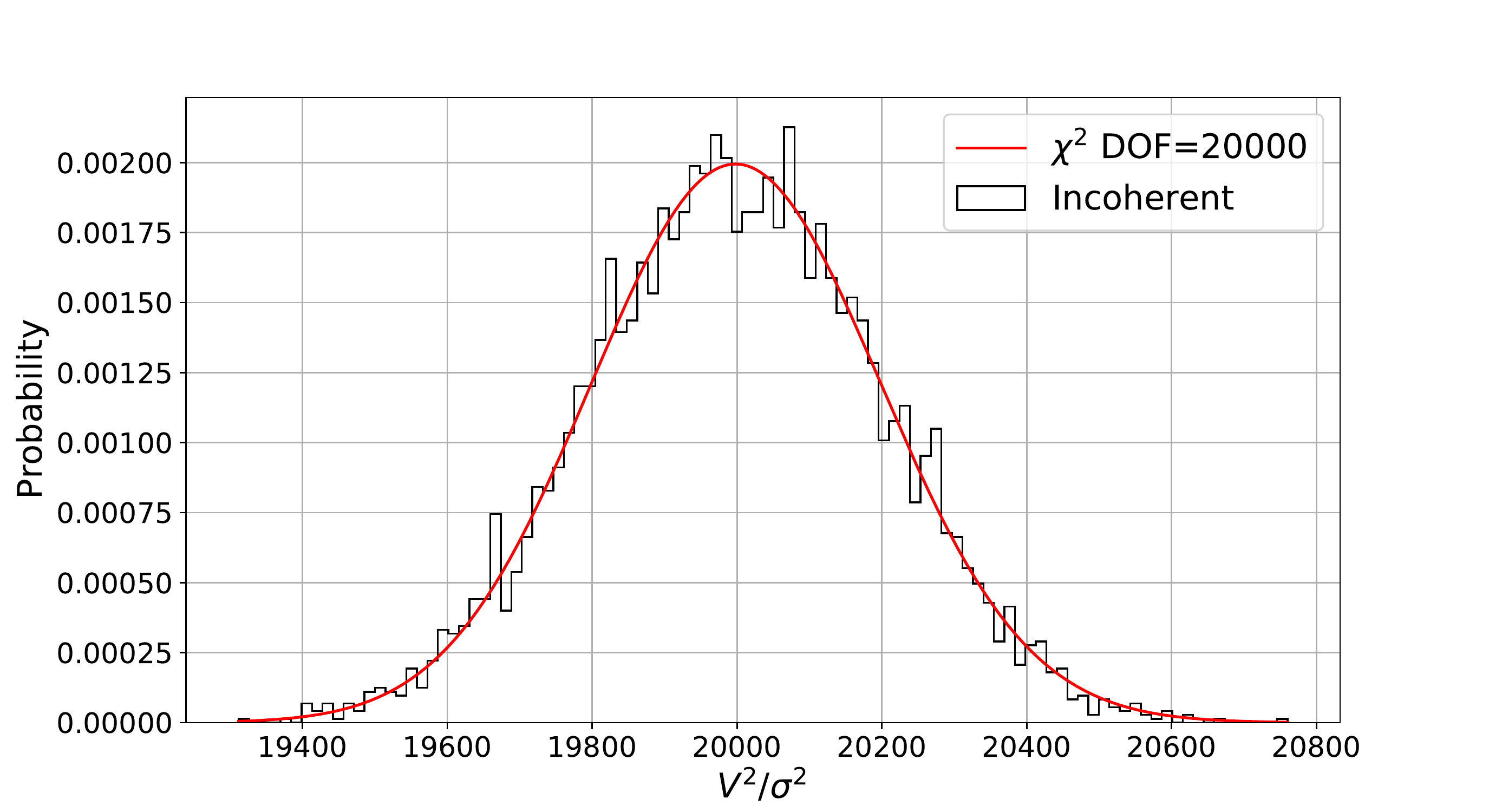}
\caption{}
\end{subfigure}
\begin{subfigure}[h]{0.49\textwidth}
\includegraphics[width=\linewidth]{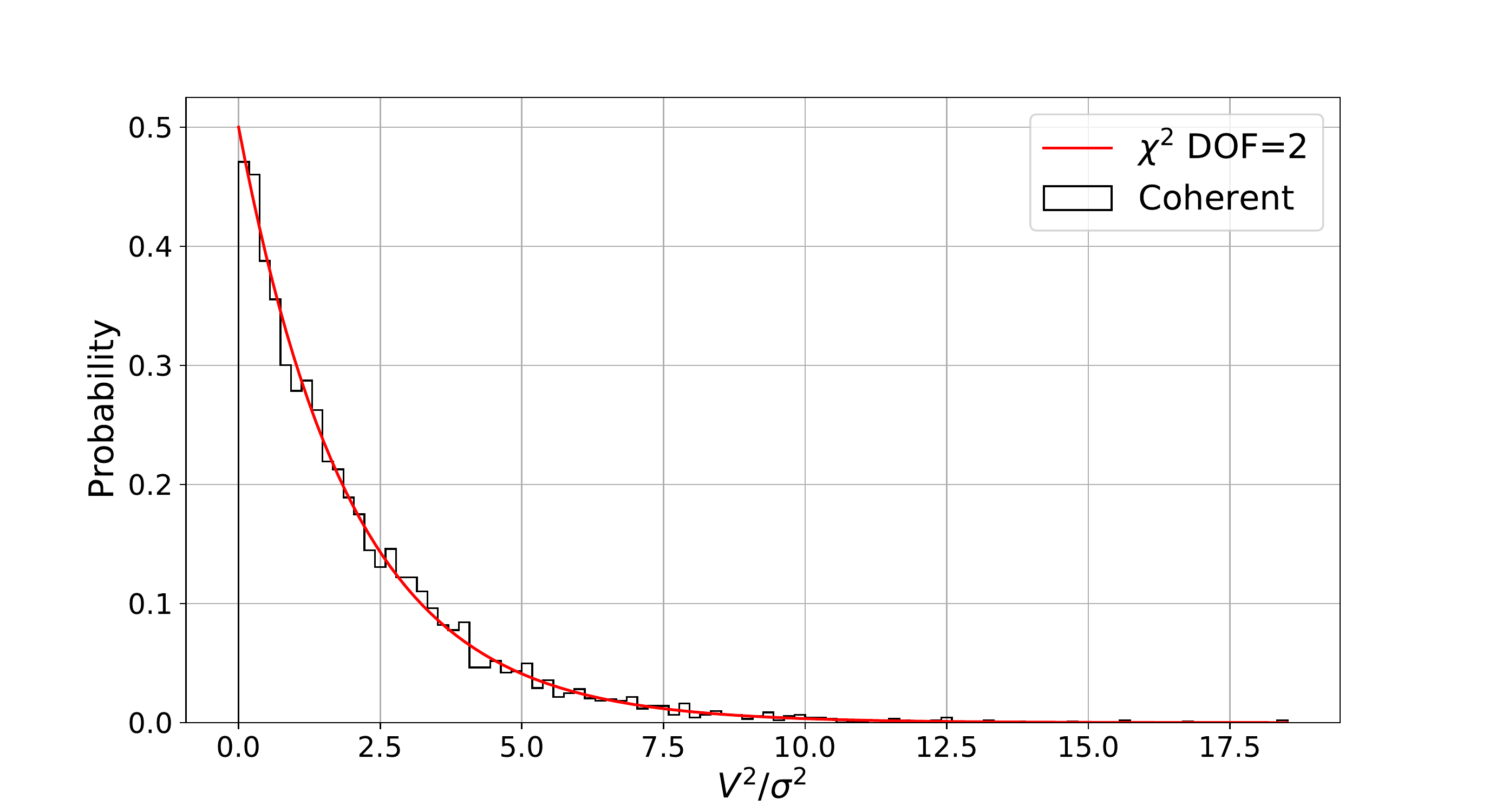}
\caption{}
\end{subfigure}
\caption{Histogram of the normalized power spectrum distribution with incoherent and coherent averaging methods. }
\label{fig histogram_sim}
\end{figure}

These estimations can be identified by numerical calculation.  We generate white noise with power spectral density $0.2\,\rm{V^{2}/Hz}$ with a sampling rate 10\,kHz. The time interval for a single time series $x_{i}$ is 1 second. Total 10000-time series are generated with the same density and sampling rate. The coherent averaging and incoherent averaging are performed for this randomly generated time-series data. Figure~\ref{fig histogram_sim} shows the numerically calculated distributions with certain normalization constant and corresponding chi-square distribution.
The probability distribution of incoherent averaging follows the chi-square distribution with the degree of freedom $2\times 10^{4}$, as we expected. That of coherent average follows the chi-square distribution of the degree of freedom 2.

From the derived distribution, the SNR relation that we numerically calculated in the previous section can be derived explicitly. The $\chi^{2}_{k}$ distribution has a mean value $k$ and a variance $2k$. The normalization constant for the power spectrum density is labeled as $\alpha\equiv 2\sigma/LN$, then SNR for both averaging methods are calculated as:
\begin{equation}
\begin{split}
&\mathrm{SNR_{p}^{i}}=\frac{X_{0}/\alpha-2N}{\sqrt{4N}}\\
&\mathrm{SNR_{p}^{c}}=\frac{X_{0}/\alpha-2}{\sqrt{4}},
\end{split}
\end{equation}
where $X_{0}$ is power spectrum of the signal. The normalization constant $\alpha\propto 1/N$, therefore the SNR has following relationship:
 \begin{equation}
\begin{split}
&\mathrm{SNR_{p}^{i}}=\left(\frac{\beta}{2}-1\right)\sqrt{N}\\
&\mathrm{SNR_{p}^{c}}=\left(\beta-\frac{1}{N}\right)N,
\end{split}
\end{equation}
where $\beta=X_{0}L/4\sigma$. The relationship of SNR to the repetition number N follows expected result. Clearly, for $1\leq N$, the coherent averaging method has always have higher SNR than that of incoherent averaging.

\bibliographystyle{utphys}
\bibliography{references}

\end{document}